\documentclass[twocolumn,pre]{revtex4}
\usepackage{graphicx}
\usepackage{amsmath}
\usepackage{amsfonts}
\usepackage{mathtools}
\usepackage{psfrag}
\usepackage[utf8]{inputenc}

\begin{document}
\title{Universal Bound on the Fano Factor in Enzyme Kinetics}

\author{Andre C. Barato and Udo Seifert}
\affiliation{ II. Institut f\"ur Theoretische Physik, Universit\"at Stuttgart, 70550 Stuttgart, Germany}

\parskip 1mm
\def\d{{\rm d}}
\def\Ps{{P_{\scriptscriptstyle \hspace{-0.3mm} s}}}
\def\MF{{\mbox{\tiny \rm \hspace{-0.3mm} MF}}}
\def\i{\text{\scriptsize $\cal{I}$}} 
\def\A{\mathcal{A}}

\begin{abstract}
The Fano factor, an observable quantifying fluctuations of product generation by a single enzyme, can reveal information about the 
underlying reaction scheme. A lower bound on this Fano factor that depends on the thermodynamic affinity
driving the transformation from substrate to product constrains the number of intermediate states of an enzymatic cycle. So far, this bound 
has been proven only for a unicyclic network of states. We show that the bound can be extended to arbitrary multicyclic networks,
with the Fano factor constraining the largest value of the effective length, which is the ratio between the number of states and 
the number of products, among all cycles.

\end{abstract}
\maketitle
%%%%%%%%%%%%%%%%%%%%%%%%%%%%%%%%%%%%%%%%%%%%%%%%%%%%%%%%%%%%%%%%%%%%%%%%%%%%%%%%%%%%%%%%%%%%%%%%%%%%%%%%%%%%%%%%%%%%%%%%%%%%%%%%%%%%%
%%%%%%%%%%%%%%%%%%%%%%%%%%%%%%%%%%%%%%%%%%%%%%%%%%%%%%%%%%%%%%%%%%%%%%%%%%%%%%%%%%%%%%%%%%%%%%%%%%%%%%%%%%%%%%%%%%%%%%%%%%%%%%%%%%%%%
\section{Introduction}
%%%%%%%%%%%%%%%%%%%%%%%%%%%%%%%%%%%%%%%%%%%%%%%%%%%%%%%%%%%%%%%%%%%%%%%%%%%%%%%%%%%%%%%%%%%%%%%%%%%%%%%%%%%%%%%%%%%%%%%%%%%%%%%%%%%%%
%%%%%%%%%%%%%%%%%%%%%%%%%%%%%%%%%%%%%%%%%%%%%%%%%%%%%%%%%%%%%%%%%%%%%%%%%%%%%%%%%%%%%%%%%%%%%%%%%%%%%%%%%%%%%%%%%%%%%%%%%%%%%%%%%%%%%

The Michaelis-Menten (MM) expression for the rate of product formation in terms of the substrate concentration
is a central paradigm in enzymatic kinetics \cite{corn13}. In this case, the reaction scheme is very simple, with one
intermediate state, corresponding to a compound of enzyme and substrate, and the further assumption that the 
generation of product is irreversible.

However, enzymatic reactions schemes are generally more complex. They can involve several intermediate states and, possibly, different enzymatic cycles,
arising, for example, from different conformational forms of the enzyme. Single molecule experiments \cite{svob93,rito06,gree07,corn07,moff08} yield precise data on the dynamics of single
enzymes, including fluctuations related to the rate of product formation. Obtaining information about the underlying enzymatic network 
through the measurement of fluctuations constitutes a promising field known as statistical kinetics \cite{schn95,shae05,moff10,moff14}.

One important quantity characterizing the dispersion of a probability distribution is the Fano factor \cite{moff14,rond09,chau13,chau14},
which is the variance squared divided by the mean. For unicyclic enzymatic networks the Fano factor related to the fluctuating product creation is
bounded from below by the inverse of the number of intermediate states \cite{koza02,kolo07,moff14}. Hence, 
its measurement provides information on the minimal number of intermediate states in the enzymatic cycle.

This bound on the Fano factor does not take into account the affinity driving the enzymatic process, which is given
by the chemical potential difference between substrate and product. Often the chemical reaction leading to product generation is assumed
to be irreversible, as is the case of the MM scheme, which would lead to a formally divergent affinity. Considering only thermodynamically
consistent models of enzymatic kinetics, with no irreversible transitions, we have recently shown that
a stronger bound on the Fano factor that depends on the affinity can be obtained \cite{bara14}. 

Proofs of these bounds assume a unicyclic network of states \cite{koza02,bara14}. However, chemical reaction networks can involve many different 
cycles. In this paper we obtain an affinity dependent bound for arbitrary multicyclic networks of states. 
Calculating the Fano factor for different multicyclic networks exactly we observe that a version of the bound for the unicyclic case  can be adapted 
to general networks. 

The paper is organized as follows. In section \ref{sec2}, we state the bound on the Fano factor, which can be proved for unicyclic networks.
In section \ref{sec3}, we infer from analyzing two case studies that this bound can be adapted  to multicyclic networks. 
The constraints imposed by measurements of the Fano factor on network topology for the case where the affinity is known are discussed
in section \ref{sec4}. We conclude in section \ref{sec5}. The method used to calculate the Fano factor is explained in Appendix \ref{appA}.
In Appendix \ref{appBnew} we provide explict calculations for a simple example. Appendix \ref{appB} contains two further examples of multicyclic networks demonstrating the validity of the bound.    

%%%%%%%%%%%%%%%%%%%%%%%%%%%%%%%%%%%%%%%%%%%%%%%%%%%%%%%%%%%%%%%%%%%%%%%%%%%%%%%%%%%%%%%%%%%%%%%%%%%%%%%%%%%%%%%%%%%%%%%%%%%%%%%%%%%%%
%%%%%%%%%%%%%%%%%%%%%%%%%%%%%%%%%%%%%%%%%%%%%%%%%%%%%%%%%%%%%%%%%%%%%%%%%%%%%%%%%%%%%%%%%%%%%%%%%%%%%%%%%%%%%%%%%%%%%%%%%%%%%%%%%%%%%
\section{Unicyclic networks}
\label{sec2}
%%%%%%%%%%%%%%%%%%%%%%%%%%%%%%%%%%%%%%%%%%%%%%%%%%%%%%%%%%%%%%%%%%%%%%%%%%%%%%%%%%%%%%%%%%%%%%%%%%%%%%%%%%%%%%%%%%%%%%%%%%%%%%%%%%%%%
%%%%%%%%%%%%%%%%%%%%%%%%%%%%%%%%%%%%%%%%%%%%%%%%%%%%%%%%%%%%%%%%%%%%%%%%%%%%%%%%%%%%%%%%%%%%%%%%%%%%%%%%%%%%%%%%%%%%%%%%%%%%%%%%%%%%%

%%%%%%%%%%%%%%%%%%%%%%%%%%%%%%%%%%%%%%%%%%%%%%%%%%%%%%%%%%%%%%%%%%%%%%%%%%%%%%%%%%%%%%%%%%%%%%%%%%%%%%%%%%%%%%%%%%%%%%%%%%%%%%%%%%%%%
\subsection{MM kinetics}
%%%%%%%%%%%%%%%%%%%%%%%%%%%%%%%%%%%%%%%%%%%%%%%%%%%%%%%%%%%%%%%%%%%%%%%%%%%%%%%%%%%%%%%%%%%%%%%%%%%%%%%%%%%%%%%%%%%%%%%%%%%%%%%%%%%%%

A standard example of a unicyclic network of states is provided by MM kinetics with reversible product generation, which is represented as
\begin{equation}
 E+S\xrightleftharpoons[k_1^-]{k_1^+} ES \xrightleftharpoons[k_2^-]{k_2^+} E+P,
\label{MMreaction} 
\end{equation}
where $E$ is the enzyme, $S$ a substrate molecule and $P$ a product molecule \cite{corn13}. The transition rate
$k_1^+$ is proportional to substrate concentration $[S]$, whereas the transition rate $k_2^-$
is proportional to product concentration $[P]$. It is assumed that $[S]$ and $[P]$ are large enough
so that they remain approximately constant. Moreover, the difference in chemical potentials $\mu_s>\mu_p$ drives
the formation of product creation with an affinity 
\begin{equation}
\A= \mu_s-\mu_p.
\label{affchem}
\end{equation}
In terms of the transition rates,
the affinity is given by the generalized detailed balance condition \cite{qian07,seif12}
\begin{equation}
\A/(k_BT)= \ln[k_1^+k_2^+/(k_1^-k_2^-)],
\end{equation}
where $T$ is the temperature and $k_B$ Boltzmann's constant. In the following, we set $k_B T= 1$, so that the affinity and chemical potentials are measured in units of $k_BT$.    

For a single enzyme, the time evolution of the model represented in Eq. (\ref{MMreaction}) is determined by the master equation, which reads
\begin{equation}
\frac{d}{dt}P_1(t)= (k_2^++k_1^-)P_2(t)-(k_1^++k_2^-)P_1(t),
\end{equation}
where $P_1(t)$ is the probability that the enzyme is free, i.e. in state $E$, at time $t$ and $P_2(t)= 1-P_1(t)$ is the probability of state $ES$, with the enzyme bound by a substrate. 
Independent of initial conditions, the stationary probability 
is reached for large $t$ and given by $P_1= (k_2^++k_1^-)/(k_1^-+k_1^++k_2^-+k_2^+)$, which is obtained by setting $\frac{d}{dt}P_1=0$.
In this stationary state, the rate at which the product $P$ is generated is given by 
\begin{equation}
V= (1-P_1) k_2^+-P_1k_2^-= \frac{k_1^+k_2^+-k_1^-k_2^-}{k_1^++k_2^++k_1^-+k_2^-}.
\label{eqeqV}
\end{equation}  

We are interested in a random variable $X$ that counts the number of generated product molecules. Moreover, we consider that $t$ is large enough so that 
\begin{equation}
\langle X\rangle= V t,
\end{equation}
where the brackets denote an average over fluctuating realizations. The Fano factor is defined as
\begin{equation}
F= \frac{\langle X^2\rangle-\langle X\rangle^2}{\langle X\rangle}= 2 \frac{D}{V},
\label{defFano}
\end{equation}
where 
\begin{equation}
D= (\langle X^2\rangle-\langle X\rangle^2)/(2t)
\end{equation}
is known as dispersion. Whereas the rate of product creation $V$ has a simple expression in terms of the stationary probability, 
this is, in general, not the case for the dispersion $D$ \cite{derr83}. For the model in Eq. (\ref{MMreaction}) the dispersion becomes \cite{derr83,koza99}
\begin{equation}
D= \frac{k_1^+k_2^++k_1^-k_2^--2V^2}{2(k_1^++k_2^++k_1^-+k_2^-)},
\end{equation}
where $V$ is given in Eq. (\ref{eqeqV}).

%%%%%%%%%%%%%%%%%%%%%%%%%%%%%%%%%%%%%%%%%%%%%%%%%%%%%%%%%%%%%%%%%%%%%%%%%%%%%%%%%%%%%%%%%%%%%%%%%%%%%%%%%%%%%%%%%%%%%%%%%%%%%%%%%%%%%
\subsection{Bound on F for unicyclic networks}
%%%%%%%%%%%%%%%%%%%%%%%%%%%%%%%%%%%%%%%%%%%%%%%%%%%%%%%%%%%%%%%%%%%%%%%%%%%%%%%%%%%%%%%%%%%%%%%%%%%%%%%%%%%%%%%%%%%%%%%%%%%%%%%%%%%%%

We consider a general unicyclic network with $N$ states for the enzyme. For example, for the model in Eq. (\ref{MMreaction}) $N=2$. If we had 
included the intermediate state $EP$, leading to the reaction scheme
\begin{equation}
 E+S\xrightleftharpoons[k_1^-]{k_1^+} ES \xrightleftharpoons[k_2^-]{k_2^+} EP\xrightleftharpoons[k_3^-]{k_3^+} E+P,
\label{MMreaction24} 
\end{equation}
then we would get $N=3$. Furthermore, $n$ is the number of consumed substrate molecules,
which is equal to the number of  generated product molecules, in the cycle. The affinity of the cycle then becomes
\begin{equation}
n\A= \ln(\Gamma_+/\Gamma_-),
\end{equation}
where $\Gamma_+=\prod_{i=1}^N k_i^+$, $\Gamma_-=\prod_{i=1}^N k_i^-$, and $\A$ is given by Eq. (\ref{affchem}). The Fano factor 
for this general case is defined as in Eq. (\ref{defFano}).

An important bound valid for unicyclic machines is $F\ge n/N$ \cite{koza02,kolo07,moff14}. Hence, measurements of the Fano factor provide a bound on the number of intermediate states $N$. 
Recently, we have obtained a stronger, affinity dependent bound on $F$ given by \cite{bara14}
\begin{equation}
F\ge\frac{n}{N}\coth\left(\frac{n\A }{2N}\right)\ge n/N. 
\label{mainbound}
\end{equation}
This improved bound becomes $F\ge n/N$ if $\A$ is large. An important point for the discussion in the next section is that this bound is saturated if 
the rates are uniform, i.e., $k_i^+=\Gamma_+^{1/N}$ and $k_i^-=\Gamma_-^{1/N}$, for $i=1,\ldots,N$ \cite{bara14}.

%%%%%%%%%%%%%%%%%%%%%%%%%%%%%%%%%%%%%%%%%%%%%%%%%%%%%%%%%%%%%%%%%%%%%%%%%%%%%%%%%%%%%%%%%%%%%%%%%%%%%%%%%%%%%%%%%%%%%%%%%%%%%%%%%%%%%
%%%%%%%%%%%%%%%%%%%%%%%%%%%%%%%%%%%%%%%%%%%%%%%%%%%%%%%%%%%%%%%%%%%%%%%%%%%%%%%%%%%%%%%%%%%%%%%%%%%%%%%%%%%%%%%%%%%%%%%%%%%%%%%%%%%%%
\section{Multicyclic networks}
%%%%%%%%%%%%%%%%%%%%%%%%%%%%%%%%%%%%%%%%%%%%%%%%%%%%%%%%%%%%%%%%%%%%%%%%%%%%%%%%%%%%%%%%%%%%%%%%%%%%%%%%%%%%%%%%%%%%%%%%%%%%%%%%%%%%%
%%%%%%%%%%%%%%%%%%%%%%%%%%%%%%%%%%%%%%%%%%%%%%%%%%%%%%%%%%%%%%%%%%%%%%%%%%%%%%%%%%%%%%%%%%%%%%%%%%%%%%%%%%%%%%%%%%%%%%%%%%%%%%%%%%%%%
\label{sec3}

Enzymatic networks can be more complex than a single cycle. In this section we address the question of 
how the bound (\ref{mainbound}) can be adapted to arbitrary multicyclic networks.

%%%%%%%%%%%%%%%%%%%%%%%%%%%%%%%%%%%%%%%%%%%%%%%%%%%%%%%%%%%%%%%%%%%%%%%%%%%%%%%%%%%%%%%%%%%%%%%%%%%%%%%%%%%%%%%%%%%%%%%%%%%%%%%%%%%%%
\subsection{First case study}
%%%%%%%%%%%%%%%%%%%%%%%%%%%%%%%%%%%%%%%%%%%%%%%%%%%%%%%%%%%%%%%%%%%%%%%%%%%%%%%%%%%%%%%%%%%%%%%%%%%%%%%%%%%%%%%%%%%%%%%%%%%%%%%%%%%%%

%%%%%%%%%%%%%%%%%%%%%%%%%%%%%%%%%%%%%%%%%%%%%%%%%%%%%%%%%%%%%%%%%%%%%%%%%%%%%%%%%%%%%%%%%%%%%%%%%%%%%%%%%%%%%%%%%%%%%%%%%%%%%%%%%%%%%
\begin{figure}
\includegraphics[width=50mm]{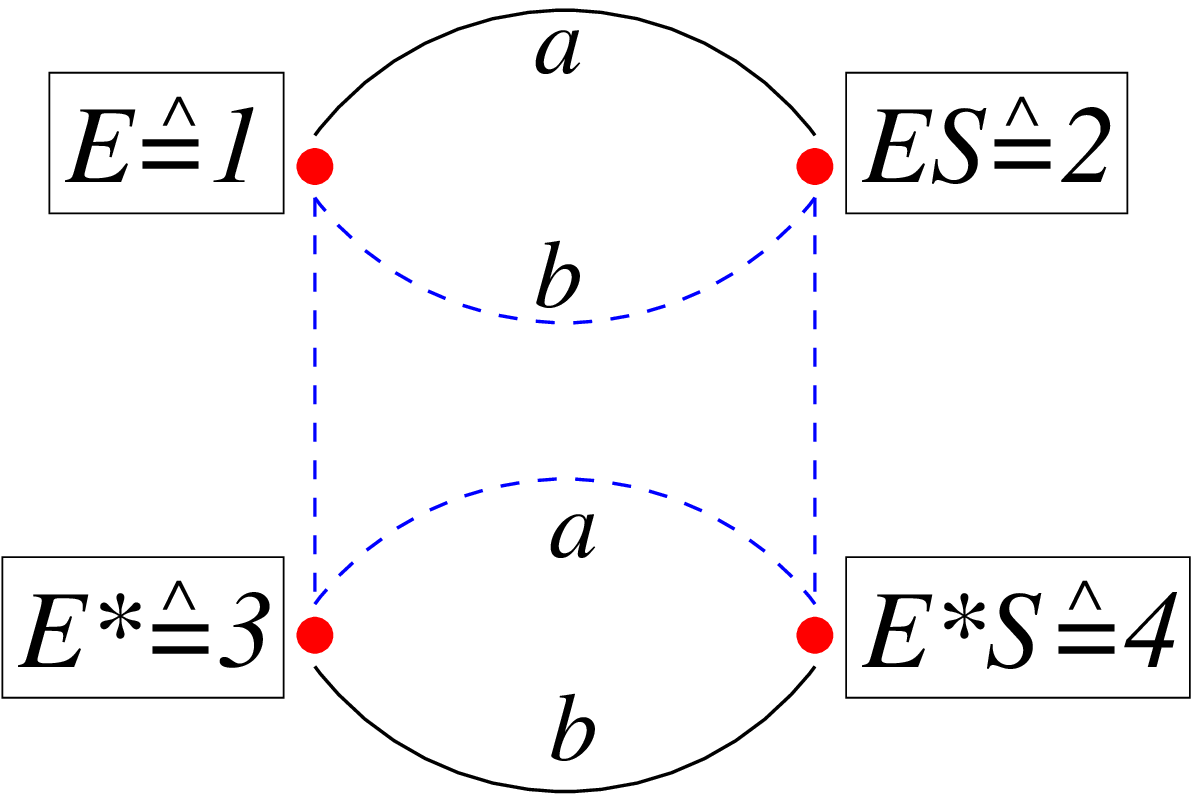}
\vspace{-2mm}
\caption{Enzymatic network for the four-state model. The links represented by $a$ are related to chemical reactions involving substrate $S$
and the links represented by $b$ are related to chemical reactions involving product $P$, in agreement with Eqs. (\ref{MMreactionnew}) and (\ref{MMreaction2}).
The dashed links form a four-state cycle that leads to the minimal Fano factor $F$.
}
\label{fig1} 
\end{figure}
%%%%%%%%%%%%%%%%%%%%%%%%%%%%%%%%%%%%%%%%%%%%%%%%%%%%%%%%%%%%%%%%%%%%%%%%%%%%%%%%%%%%%%%%%%%%%%%%%%%%%%%%%%%%%%%%%%%%%%%%%%%%%%%%%%%%%

As a first example of a multicyclic network of states we consider an enzyme $E$ that can also be in a second conformational state $E^*$, see Fig. \ref{fig1}. 
The effect of having such conformational states on the MM expression has been recently analyzed in \cite{cao11,kolo11}.
There are four possible states, which are identified as $E\mathrel{\hat=} 1$, $ES\mathrel{\hat=} 2$, $E^*\mathrel{\hat=} 3$ and $E^*S\mathrel{\hat=} 4$.
Besides the reaction scheme in Eq. (\ref{MMreaction}), which now takes place with rates
\begin{equation}
 E+S\xrightleftharpoons[k_{21}^a]{k_{12}^a} ES \xrightleftharpoons[k_{12}^b]{k_{21}^b} E+P,
\label{MMreactionnew} 
\end{equation}
there are also the reactions
\begin{equation}
E^*+S\xrightleftharpoons[k_{43}^a]{k_{34}^a} E^*S \xrightleftharpoons[k_{34}^b]{k_{43}^b} E^*+P,
\label{MMreaction2} 
\end{equation}
\begin{equation}
E\xrightleftharpoons[k_{31}]{k_{13}} E^*,
\label{reaction1E} 
\end{equation}
and
\begin{equation}
ES\xrightleftharpoons[k_{42}]{k_{24}} E^*S.
\label{reaction1ES} 
\end{equation}
The enzymatic network corresponding to these chemical reactions is shown in Fig. \ref{fig1}. 

The transition rates for this model are not all independent. For thermodynamic consistency, the following constraints must be fulfilled.
Since the affinity of cycles where one $S$ is consumed and one $P$ generated is $\A$, from the cycle in Eq. (\ref{MMreactionnew}) we get
\begin{equation}
\ln\frac{k_{12}^ak_{21}^b}{k_{21}^ak_{12}^b}= \A,
\end{equation} 
and, from the cycle in Eq. (\ref{MMreaction2}) we obtain 
\begin{equation}
\ln\frac{k_{34}^ak_{43}^b}{k_{43}^ak_{34}^b}= \A.
\end{equation} 
Moreover, the cycle 
\begin{equation}
E+S\xrightleftharpoons[k_{21}^a]{k_{12}^a} ES \xrightleftharpoons[k_{42}]{k_{24}} E^*S \xrightleftharpoons[k_{34}^a]{k_{43}^a} E^*+S \xrightleftharpoons[k_{13}]{k_{31}} E+S,
\end{equation}
which does not lead to product formation, has affinity $0$, leading to the constraint
\begin{equation}
\ln\frac{k_{12}^ak_{24}k_{43}^ak_{31}}{k_{21}^ak_{42}k_{34}^ak_{13}}= 1.
\end{equation}
Hence, these $3$ constraints for $12$ rates lead to $9$ independent transition rates.

Denoting the probability of being in state $i$ at time $t$ by $P_i(t)$, the master equation reads
\begin{equation}
\frac{d}{dt}\mathbf{P}(t)=  \mathbf{L}\mathbf{P}(t)
\label{mastereq}
\end{equation}
where $\mathbf{P}(t)=\{P_1(t),P_2(t),P_3(t),P_4(t)\}$ and the stochastic matrix $\mathbf{L}$
is
\begin{equation}
\left(
\begin{array}{cccc}
-r_1 & k^a_{21}+k^b_{21} & k_{31} & 0 \\
 k^a_{12}+k^b_{12}          & -r_2 & 0 & k_{42} \\ 
 k_{13}          & 0 & -r_3 & k^a_{43}+k^b_{43} \\
0          & k_{24} & k^a_{34}+k^b_{34} & -r_4
\end{array}
\right),  
\label{matrix1}
\end{equation}
where $r_1= k^a_{12}+k^b_{12}+k_{13}$, $r_2= k^a_{21}+k^b_{21}+k_{24}$, $r_3= k_{31}+k^a_{34}+k^b_{34}$, and $r_4= k^a_{43}+k^b_{43}+k_{42}$.
The stationary probability vector $\mathbf{P}$ is obtained from  $\mathbf{L}\mathbf{P}=0$. Furthermore, the rate of product generation
is $V= P_2 k_{21}^b-P_1k_{12}^b+P_4 k_{43}^b-P_3k_{34}^b$. The calculation of the dispersion $D$, as explained in Appendix \ref{appA}, leads to a quite long expression
for the Fano factor $F$ as a function of the affinity $\A$ and $9$ independent transition rates. 

%%%%%%%%%%%%%%%%%%%%%%%%%%%%%%%%%%%%%%%%%%%%%%%%%%%%%%%%%%%%%%%%%%%%%%%%%%%%%%%%%%%%%%%%%%%%%%%%%%%%%%%%%%%%%%%%%%%%%%%%%%%%%%%%%%%%%
\begin{figure}
\psfrag{A}{$\mathcal{A}$}
\includegraphics[width=75mm]{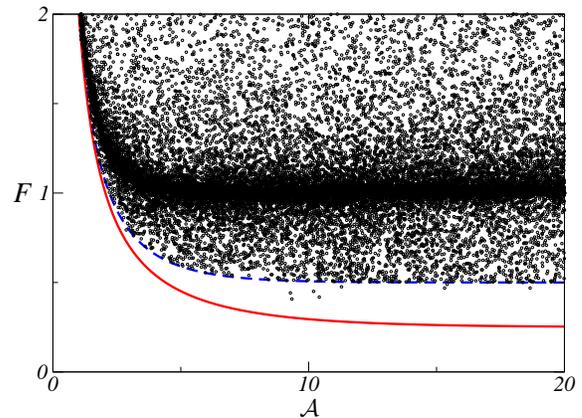}
\vspace{-2mm}
\caption{The Fano factor $F$ as a function of the chemical potential difference $\A$, evaluated at randomly chosen rates, for the model shown in Figure \ref{fig1}.
The red solid line corresponds to the bound $(1/4)\coth(\A/8)$.
The transition rates $k_{21}^b$, $k_{43}^b$, and $k_{13}$ are eliminated with the relations $k_{21}^b= \textrm{e}^\A k_{12}^bk_{21}^a/k_{12}^a$,  $k_{43}^b= \textrm{e}^\A k_{34}^bk_{43}^a/k_{34}^a$,
and $k_{13}= k_{12}^ak_{24}k_{43}^ak_{31}/(k_{21}^ak_{42}k_{34}^a)$. The remaining transition rates are chosen by generating a random number $y$ between $-5$ and $5$ and then taking $k=10^y$.
This symmetric choice of the rates lead to the property that only a few points go below $(1/2)\coth(\A/4)$, which is shown as a blue dashed line, indicating that the region in the 
space of transition rates where $F$ is close to the bound is small.
}
\label{fig2} 
\end{figure}
%%%%%%%%%%%%%%%%%%%%%%%%%%%%%%%%%%%%%%%%%%%%%%%%%%%%%%%%%%%%%%%%%%%%%%%%%%%%%%%%%%%%%%%%%%%%%%%%%%%%%%%%%%%%%%%%%%%%%%%%%%%%%%%%%%%%%

In Fig. \ref{fig2}, where we evaluate the Fano factor at randomly chosen transition rates obeying the constraints, we see that $F$ is bounded by the right hand side of Eq. (\ref{mainbound}) with $N=4$ and $n=1$.
This bound is related to cycles in the network of Fig. \ref{fig2} which have $N=4$ states. An example of such a cycle, as represented by the dashed links in Fig. \ref{fig1}, is 
\begin{equation}
E^*+S\xrightleftharpoons[k_{43}^a]{k_{34}^a} E^*S \xrightleftharpoons[k_{24}]{k_{42}} ES \xrightleftharpoons[k_{12}^b]{k_{21}^b} E+P\xrightleftharpoons[k_{31}]{k_{13}} E^*+P.
\end{equation}
Choosing uniform transition rates for this cycle and considering the transition rates that are not part of the cycle to be small, $F$ should indeed reach the bound in Eq. (\ref{mainbound}) with $N=4$. For example,
setting $k_{34}^a=k_{42}=k_{21}^b=k_{13}= \exp(\A/4)$, $k_{43}^a=k_{24}=k_{12}^b=k_{31}= 1$,  $k_{12}^a=k_{43}^b= k\exp(3\A/4)$, and $k_{21}^a=k_{34}^b= k$ in the expression for $F$, by taking the limit $k\to 0$
we obtain analytically 
\begin{equation}
F= \frac{1}{4}\coth\left(\frac{\A}{8}\right).
\end{equation}
In Appendix \ref{appBnew}, we consider a simpler network, where a similar calculation can be performed explicitly.  
As an independent check that the Fano factor is indeed larger than the above expression we have also minimized it numerically.
Hence, even though the MM reaction scheme in Eqs. (\ref{MMreactionnew}) and (\ref{MMreaction2}) has two states, the Fano factor for the full model in Fig. \ref{fig1} can be smaller than the bound in Eq. (\ref{mainbound}) with $N=2$:
the conformational changes in the enzyme allow for larger cycles with $N=4$. The bound in Eq. (\ref{mainbound}) for unicyclic networks thus also applies for the multicyclic network of Fig. \ref{fig1}, 
with $N$ being the size of the largest cycle in the network. 
 
%%%%%%%%%%%%%%%%%%%%%%%%%%%%%%%%%%%%%%%%%%%%%%%%%%%%%%%%%%%%%%%%%%%%%%%%%%%%%%%%%%%%%%%%%%%%%%%%%%%%%%%%%%%%%%%%%%%%%%%%%%%%%%%%%%%%%
\subsection{Second case study}
%%%%%%%%%%%%%%%%%%%%%%%%%%%%%%%%%%%%%%%%%%%%%%%%%%%%%%%%%%%%%%%%%%%%%%%%%%%%%%%%%%%%%%%%%%%%%%%%%%%%%%%%%%%%%%%%%%%%%%%%%%%%%%%%%%%%%

%%%%%%%%%%%%%%%%%%%%%%%%%%%%%%%%%%%%%%%%%%%%%%%%%%%%%%%%%%%%%%%%%%%%%%%%%%%%%%%%%%%%%%%%%%%%%%%%%%%%%%%%%%%%%%%%%%%%%%%%%%%%%%%%%%%%%
\begin{figure}
\includegraphics[width=75mm]{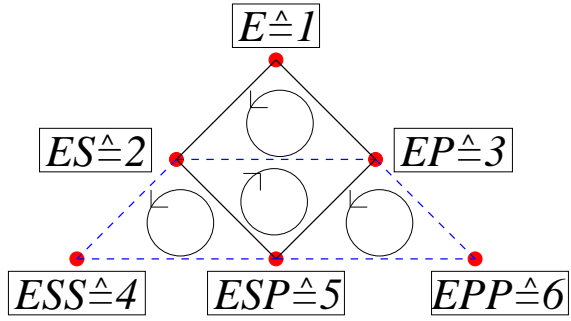}
\vspace{-2mm}
\caption{Enzymatic network for the six-state model. The dashed links are part of a five-state cycle leading to the minimal Fano factor.
}
\label{fig3} 
\end{figure}
%%%%%%%%%%%%%%%%%%%%%%%%%%%%%%%%%%%%%%%%%%%%%%%%%%%%%%%%%%%%%%%%%%%%%%%%%%%%%%%%%%%%%%%%%%%%%%%%%%%%%%%%%%%%%%%%%%%%%%%%%%%%%%%%%%%%%

As a second example of a multicyclic network we consider an enzyme that can bind two substrate molecules \cite{corn13}, as shown in Fig. \ref{fig3}. Intermediate states
with an enzyme bound with product molecules are also allowed. The model has $6$ states which are identified as   
$E\mathrel{\hat=} 1$, $ES\mathrel{\hat=} 2$, $EP\mathrel{\hat=} 3$, $ESS\mathrel{\hat=} 4$, $ESP\mathrel{\hat=} 5$, and $EPP\mathrel{\hat=} 6$.
The possible reactions are 
\begin{equation}
E+S\xrightleftharpoons[k_{21}]{k_{12}} ES \xrightleftharpoons[k_{32}]{k_{23}} EP \xrightleftharpoons[k_{13}]{k_{31}} E+P,
\end{equation}
\begin{equation}
ES+S\xrightleftharpoons[k_{42}]{k_{24}} ESS \xrightleftharpoons[k_{54}]{k_{45}} ESP \xrightleftharpoons[k_{25}]{k_{52}} ES+P,
\end{equation}
and
\begin{equation}
EP+S\xrightleftharpoons[k_{53}]{k_{35}} ESP \xrightleftharpoons[k_{65}]{k_{56}} EPP \xrightleftharpoons[k_{36}]{k_{63}} EP+P.
\end{equation}
There is also another three-state cycle, which is the middle cycle in Fig. \ref{fig3}
that does not involve further reactions.
The stochastic matrix $\mathbf{L}$, defined in Eq. (\ref{mastereq}), for this model is
\begin{equation}
\left(
\begin{array}{cccccc}
-r_1 & k_{21} & k_{31} & 0 & 0 & 0 \\
 k_{12}     & -r_2 & k_{32} & k_{42} & k_{52} & 0 \\ 
 k_{13}     & k_{23} & -r_3 & 0 & k_{53} & k_{63} \\ 
 0     & k_{24} & 0 & -r_4 & k_{54} & 0 \\ 
0    & k_{25} & k_{35} & k_{45} & -r_5 & k_{65} \\ 
 0     & 0 & k_{36} & 0 & k_{56} & -r_6 
\end{array}
\right)  
\label{matrix2}
\end{equation}
where $r_1= k_{12}+k_{13}$, $r_2= k_{21}+k_{23}+k_{24}+k_{25}$, $r_3= k_{31}+k_{32}+k_{35}+k_{36}$, $r_4= k_{42}+k_{45}$, $r_5=  k_{52}+k_{53}+k_{54}+k_{56}$,
and $r_6= k_{63}+k_{65}$. The $18$ transition rates fulfill the $4$ constraints
\begin{eqnarray}
\A & = \ln\frac{k_{12}k_{23}k_{31}}{k_{21}k_{32}k_{13}}= \ln\frac{k_{24}k_{45}k_{52}}{k_{42}k_{54}k_{25}}\nonumber\\
 &  =\ln\frac{k_{35}k_{56}k_{63}}{k_{53}k_{65}k_{36}}=\ln\frac{k_{23}k_{35}k_{52}}{k_{32}k_{53}k_{25}},
\label{eqconstraints}
\end{eqnarray}    
which come from the fact that the affinity for each of the four three-state cycles indicated in Fig. \ref{fig3} is $\A$. The rate at which product is generated can be written as 
$V= P_3 k_{31}-P_1 k_{13}+P_6 k_{63}-P_3 k_{36}+P_5 k_{52}-P_2 k_{25}$. The calculation of
the dispersion $D$, which is explained in Appendix \ref{appA}, leads to an even longer expression for $F$ as a function of $\A$ and $14$ independent transition rates. 
%%%%%%%%%%%%%%%%%%%%%%%%%%%%%%%%%%%%%%%%%%%%%%%%%%%%%%%%%%%%%%%%%%%%%%%%%%%%%%%%%%%%%%%%%%%%%%%%%%%%%%%%%%%%%%%%%%%%%%%%%%%%%%%%%%%%%
\begin{figure}
\psfrag{A}{$\mathcal{A}$}
\includegraphics[width=75mm]{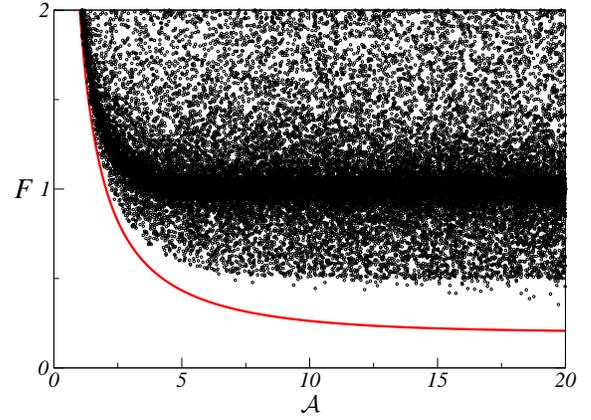}
\vspace{-2mm}
\caption{The Fano factor $F$ as a function of the chemical potential difference $\A$, evaluated at randomly chosen rates, for the model of Fig. \ref{fig3}.
The red solid line corresponds to  the bound $(1/5)\coth(\A/10)$. From the constraints in Eq. (\ref{eqconstraints}) we eliminate the transition rates $k_{31}= \textrm{e}^\A k_{21}k_{32}k_{13}/(k_{12}k_{23})$, 
$k_{63}= \textrm{e}^\A k_{36}k_{53}k_{65}/(k_{35}k_{56})$, $k_{52}= \textrm{e}^\A k_{53}k_{25}k_{32}/(k_{35}k_{23})$, and $k_{45}= k_{42}k_{54}k_{35}k_{23}/(k_{24}k_{53}k_{32})$. 
The remaining transition rates are chosen by generating a random number $y$ between $-5$ and $5$ and then taking $k=10^y$. The fact that the points are far from the bound
indicates that the region in the space of transition rates where $F$ becomes close to the bound is small.
}
\label{fig4} 
\end{figure}
%%%%%%%%%%%%%%%%%%%%%%%%%%%%%%%%%%%%%%%%%%%%%%%%%%%%%%%%%%%%%%%%%%%%%%%%%%%%%%%%%%%%%%%%%%%%%%%%%%%%%%%%%%%%%%%%%%%%%%%%%%%%%%%%%%%%%

The largest cycle for the network in Fig. \ref{fig3} is the six-state external cycle going through all states. Even though $N=6$
for this cycle, the number of consumed substrate and generated product is $n=2$. Hence, the bound on the Fano factor in Eq. (\ref{mainbound}) for this cycle with $N=6$ 
and $n=2$ is the same as the bound for the four cycles with $N=3$ and $n=1$ indicated in Fig. \ref{fig3}. There are also cycles with $N=5$ states and $n=1$,
with an example indicated by the dashed links in Fig. \ref{fig3}. The latter cycle gives the bound on the Fano factor, which is reached when the multicyclic network becomes 
effectively this cycle with uniform rates. This limit can be achieved by setting the transition rates, e.g., in the following way. For the five-state cycle we
take the uniform rates $k_{32}=k_{24}=k_{45}=k_{56}=k_{63}= \exp(\A/5)$ and $k_{23}=k_{42}=k_{54}=k_{65}=k_{36}= 1$. As the transitions related to the links $EP-ESP$ and $ES-ESP$
should be slow, we set $k_{35}=k_{52}=k\exp(3\A/5)$ and $k_{53}=k_{25}=k$, with $k$ being small. The transition to leave the cycle to state $E$ should also be slow, whereas
the transitions to return to the cycle from state $E$ should be fast, as with $k_{12}=\exp(3\A/5)/k$, $k_{21}=k$, $k_{13}=1/k$, and $k_{31}=k\exp(3\A/5)$. With this
choice for the transition rates in the expression for $F$, by taking the limit $k\to 0$, we obtain analytically
\begin{equation}
F= \frac{1}{5}\coth\left(\frac{\A}{10}\right).
\end{equation}    
We verified with numerical minimization and evaluation at randomly chosen transition rates, as illustrated in Fig. \ref{fig4}, that the Fano factor does not go below this bound.

%%%%%%%%%%%%%%%%%%%%%%%%%%%%%%%%%%%%%%%%%%%%%%%%%%%%%%%%%%%%%%%%%%%%%%%%%%%%%%%%%%%%%%%%%%%%%%%%%%%%%%%%%%%%%%%%%%%%%%%%%%%%%%%%%%%%%
\subsection{Conjecture for arbitrary networks}
%%%%%%%%%%%%%%%%%%%%%%%%%%%%%%%%%%%%%%%%%%%%%%%%%%%%%%%%%%%%%%%%%%%%%%%%%%%%%%%%%%%%%%%%%%%%%%%%%%%%%%%%%%%%%%%%%%%%%%%%%%%%%%%%%%%%%

The results obtained with the two case studies and the two further examples shown in Appendix \ref{appB} lead us to the following conjecture.
Given an enzymatic process where a substrate $S$ is consumed and product $P$ is produced with 
an arbitrary enzymatic network, the Fano factor is bounded from below by  
\begin{equation}
F\ge \frac{1}{M_{\textrm{max}}}\coth\left(\frac{\A}{2 M_{\textrm{max}}}\right),
\label{boundfinal}
\end{equation}    
where $M_{\textrm{max}}$ is the largest value of the effective length $M= N/n$, i.e., the length per product molecule, among all cycles in the multicyclic network.
A posteriori, the idea behind the conjecture is simple: a multicyclic network cannot have a Fano factor smaller
than the one of a unicyclic network with the largest effective length $M_\textrm{max}$. If the rates are chosen such that the
network is fully dominated by this cycle with uniform rates, changing the rates will add cycles with smaller $M$ leading 
to an increase in $F$. 

A somewhat related result that for arbitrary Markov processes the relative uncertainty associated with the probability 
of the time to reach an absorbing state is bounded by the inverse of the total number of states in the network has been proven rigorously \cite{aldo87}. 
This relative uncertainty is equal to the Fano factor for the special case of a unicyclic network with at least one irreversible transition \cite{moff14},
which amounts to a formally divergent affinity.     

%We emphasize that the bound $F\ge 1/M_{\textrm{max}}$, which takes place for $\A\to\infty$,
%was known to be valid for unicyclic networks: even in this case our conjecture represents an improvement.

%%%%%%%%%%%%%%%%%%%%%%%%%%%%%%%%%%%%%%%%%%%%%%%%%%%%%%%%%%%%%%%%%%%%%%%%%%%%%%%%%%%%%%%%%%%%%%%%%%%%%%%%%%%%%%%%%%%%%%%%%%%%%%%%%%%%%
%%%%%%%%%%%%%%%%%%%%%%%%%%%%%%%%%%%%%%%%%%%%%%%%%%%%%%%%%%%%%%%%%%%%%%%%%%%%%%%%%%%%%%%%%%%%%%%%%%%%%%%%%%%%%%%%%%%%%%%%%%%%%%%%%%%%%
\section{Fano factor as a diagnostic tool for network topology}
%%%%%%%%%%%%%%%%%%%%%%%%%%%%%%%%%%%%%%%%%%%%%%%%%%%%%%%%%%%%%%%%%%%%%%%%%%%%%%%%%%%%%%%%%%%%%%%%%%%%%%%%%%%%%%%%%%%%%%%%%%%%%%%%%%%%%
%%%%%%%%%%%%%%%%%%%%%%%%%%%%%%%%%%%%%%%%%%%%%%%%%%%%%%%%%%%%%%%%%%%%%%%%%%%%%%%%%%%%%%%%%%%%%%%%%%%%%%%%%%%%%%%%%%%%%%%%%%%%%%%%%%%%%
\label{sec4}
%%%%%%%%%%%%%%%%%%%%%%%%%%%%%%%%%%%%%%%%%%%%%%%%%%%%%%%%%%%%%%%%%%%%%%%%%%%%%%%%%%%%%%%%%%%%%%%%%%%%%%%%%%%%%%%%%%%%%%%%%%%%%%%%%%%%%
\begin{figure}
\psfrag{A}{$\mathcal{A}$}
\psfrag{B}{$\to\infty$}
\includegraphics[width=75mm]{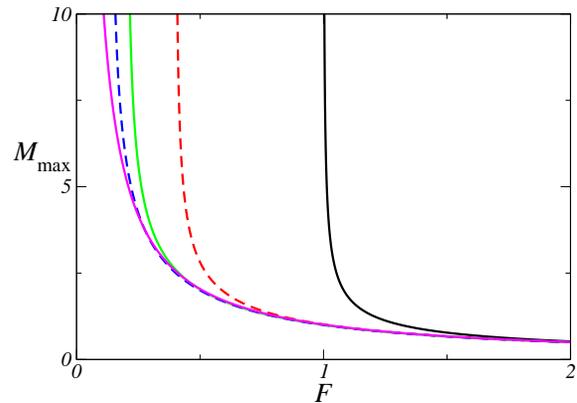}
\vspace{-2mm}
\caption{Minimal effective length of the largest cycle $M_{\textrm{max}}$, as given by Eq. (\ref{boundfinal}), as a function of the Fano factor $F$ for different affinities $\A$. 
From right to left, the values of the affinities are $\A=2$, $\A=5$, $\A=10$, $\A=15$, and $\A\to \infty$ (which corresponds to the bound $M_{\textrm{max}}\ge 1/F$).  
}
\label{fig5} 
\end{figure}
%%%%%%%%%%%%%%%%%%%%%%%%%%%%%%%%%%%%%%%%%%%%%%%%%%%%%%%%%%%%%%%%%%%%%%%%%%%%%%%%%%%%%%%%%%%%%%%%%%%%%%%%%%%%%%%%%%%%%%%%%%%%%%%%%%%%%

Our new bound (\ref{boundfinal}) can be turned into a diagnostic tool leading to some information on the topology of the underlying enzymatic network as follows.
By solving the transcendental equation (\ref{boundfinal}), we find a bound on $M_{\textrm{max}}$ that depends on the Fano factor $F$ and
the affinity $\A$, as shown in Fig. \ref{fig5}. The affinity imposes important constraints on the allowed values of $F$ and $M_{\textrm{max}}$. First, Eq. (\ref{boundfinal}) implies $F\ge 2/\A$. For example, if the 
free energy consumed when one substrate molecule is transformed into a product molecule is $2 k_B T$ (which implies $\A=2$), 
the Fano factor $F$ must be larger than $1$. Second, even far from equilibrium, i.e., for large ratio between forward and backward transitions rates, the affinity dependent bound on $M_{\textrm{max}}$ can represent a considerable 
improvement in relation to $M_{\textrm{max}}\ge 1/F$. Specifically, if the free energy consumption for transforming one molecule is $10 k_B T$, i.e., $\A=10$, a measurement that gives a Fano factor $F=0.22$ implies that the 
cycle with largest effective length has $M_{\textrm{max}}\ge 9$, which is almost twice the bound obtained for a formally divergent affinity $M_{\textrm{max}}\ge1/F\simeq 4.55$. We stress that this discussion 
is novel even for the unicyclic case. Furthermore, the size of the largest cycle in the network cannot be inferred from our bound. The cycle with $M_{\textrm{max}}$ does not need to correspond to the cycle with largest $N$, as demonstrated
in the second case study from Sec. \ref{sec3}.

%%%%%%%%%%%%%%%%%%%%%%%%%%%%%%%%%%%%%%%%%%%%%%%%%%%%%%%%%%%%%%%%%%%%%%%%%%%%%%%%%%%%%%%%%%%%%%%%%%%%%%%%%%%%%%%%%%%%%%%%%%%%%%%%%%%%%
%%%%%%%%%%%%%%%%%%%%%%%%%%%%%%%%%%%%%%%%%%%%%%%%%%%%%%%%%%%%%%%%%%%%%%%%%%%%%%%%%%%%%%%%%%%%%%%%%%%%%%%%%%%%%%%%%%%%%%%%%%%%%%%%%%%%%
\section{Conclusion}
%%%%%%%%%%%%%%%%%%%%%%%%%%%%%%%%%%%%%%%%%%%%%%%%%%%%%%%%%%%%%%%%%%%%%%%%%%%%%%%%%%%%%%%%%%%%%%%%%%%%%%%%%%%%%%%%%%%%%%%%%%%%%%%%%%%%%
%%%%%%%%%%%%%%%%%%%%%%%%%%%%%%%%%%%%%%%%%%%%%%%%%%%%%%%%%%%%%%%%%%%%%%%%%%%%%%%%%%%%%%%%%%%%%%%%%%%%%%%%%%%%%%%%%%%%%%%%%%%%%%%%%%%%%
\label{sec5}

We have found a universal lower bound on the Fano factor that takes into account the chemical potential difference involved
in the enzymatic reactions. For unicyclic networks, this result, which has been obtained in \cite{bara14}, is an improvement in relation 
to the well known bound $F\ge n/N$, which corresponds to Eq. (\ref{boundfinal}) with $\A\to \infty$. This limit of divergent affinity happens when some 
chemical reaction is assumed to be irreversible. Such an assumption typically does not do any harm if one is interested
in the rate of formation of product as a function of the substrate concentration, for example. However, in principle all
chemical reactions are reversible and thermodynamic consistent models must have a finite affinity. 

Based on several case studies supporting the idea that a multycyclic network does not lead to a Fano factor smaller
than the bound associated with the cycle with largest effective length $M=N/n$, we here conjecture that this bound also 
holds for arbitrary enzymatic schemes, with a possibly complex multicyclic network of states. 
Measurements of the Fano factor will now provide information on the size of the cycle with largest effective length, independent of the 
details of the underlying enzymatic network. Provided the affinity is known, as, e.g., in a recent experiment on the F1-ATPase \cite{toya11},
our bound can represent a considerable improvement in relation to $M_{\textrm{max}}\ge 1/F$, as represented in Fig. \ref{fig5}.

We have restricted our analysis to enzymatic processes for which there is only one chemical potential difference. 
Investigating possible bounds on the Fano factor for the case where different substrates and products
lead to several chemical potential differences would be interesting. Finally, on a technical side, 
we stress that finding an algebraic proof of our well-supported conjecture for arbitrary enzymatic networks represents 
a serious mathematical challenge.

%\begin{acknowledgments}
%\end{acknowledgments}

%==========================================================================
% References
%==========================================================================
\appendix

%%%%%%%%%%%%%%%%%%%%%%%%%%%%%%%%%%%%%%%%%%%%%%%%%%%%%%%%%%%%%%%%%%%%%%%%%%%%%%%%%%%%%%%%%%%%%%%%%%%%%%%%%%%%%%%%%%%%%%%%%%%%%%%%%%%%%
%%%%%%%%%%%%%%%%%%%%%%%%%%%%%%%%%%%%%%%%%%%%%%%%%%%%%%%%%%%%%%%%%%%%%%%%%%%%%%%%%%%%%%%%%%%%%%%%%%%%%%%%%%%%%%%%%%%%%%%%%%%%%%%%%%%%%
\section{Expression for the dispersion $D$}
%%%%%%%%%%%%%%%%%%%%%%%%%%%%%%%%%%%%%%%%%%%%%%%%%%%%%%%%%%%%%%%%%%%%%%%%%%%%%%%%%%%%%%%%%%%%%%%%%%%%%%%%%%%%%%%%%%%%%%%%%%%%%%%%%%%%%
%%%%%%%%%%%%%%%%%%%%%%%%%%%%%%%%%%%%%%%%%%%%%%%%%%%%%%%%%%%%%%%%%%%%%%%%%%%%%%%%%%%%%%%%%%%%%%%%%%%%%%%%%%%%%%%%%%%%%%%%%%%%%%%%%%%%%
\label{appA}
In this Appendix we explain the method we use to calculate the Fano factor, which is due to Koza \cite{koza99}. 
We consider a general network of states with transition rates from $i$ to $j$ denoted by $k_{ij}$. The total number of states
in the network is $Q$. We define a $Q\times Q$ matrix, which is a function of a new variable $z$, as  
\begin{equation}	
[\mathbf{L}(z)]_{ji}=\left\{\begin{array}{ll} 
 k_{ij}\exp(z d_{ij}) & \quad \textrm{if } i\neq j\\
 -\sum_jk_{ij} & \quad \textrm{if } i=j
\end{array}\right.\,.
\label{modgenA}
\end{equation}
The generalized distance $d_{ij}$ characterizes how much the random variable $X$ changes in a transition from $i$ to $j$. If a transition from $i$ to $j$ involves the enzyme liberating a product molecule
then $d_{ij}=1$ and $d_{ji}=-1$. If a transition between $i$ and $j$ does not involve a product molecule, then $d_{ij}=0$. As explicit examples we consider the two case studies analyzed
in section \ref{sec3}. For the model in Fig. \ref{fig1}, $\mathbf{L}(z)$ becomes
\begin{equation}
\left(
\begin{array}{cccc}
-r_1 & k^a_{21}+l^b_{21} & k_{31} & 0 \\
 k^a_{12}+l^b_{12}          & -r_2 & 0 & k_{42} \\ 
 k_{13}          & 0 & -r_3 & k^a_{43}+l^b_{43} \\
0          & k_{24} & k^a_{34}+l^b_{34} & -r_4
\end{array}
\right),  
\end{equation}
where $l_{12}^b=k_{12}^b\textrm{e}^{-z}$, $l_{34}^b=k_{34}^b\textrm{e}^{-z}$, $l_{21}^b=k_{21}^b\textrm{e}^{z}$, $l_{43}^b=k_{43}^b\textrm{e}^{z}$. 
For the model in Fig. \ref{fig3}, $\mathbf{L}(z)$  is given by
\begin{equation}
\left(
\begin{array}{cccccc}
-r_1 & k_{21}  & k_{31}\textrm{e}^z & 0 & 0 & 0 \\
 k_{12}     & -r_2 & k_{32} & k_{42} & k_{52}\textrm{e}^{z} & 0 \\ 
 k_{13}\textrm{e}^{-z}    & k_{23} & -r_3 & 0 & k_{53} & k_{63}\textrm{e}^{z} \\ 
 0     & k_{24} & 0 & -r_4 & k_{54} & 0 \\ 
0    & k_{25}\textrm{e}^{-z} & k_{35} & k_{45} & -r_5 & k_{65} \\ 
 0     & 0 & k_{36} \textrm{e}^{-z} & 0 & k_{56} & -r_6 
\end{array}
\right).  
\end{equation}

It can be shown that the maximum eigenvalue of $\mathbf{L}(z)$, which we denote by $\lambda(z)$, is the scaled cumulant generating function associated
with the random variable $X$ \cite{lebo99}. Hence, the following expressions are valid   
\begin{equation}
V= \lambda'
\label{Vcum}
\end{equation}
and
\begin{equation}
D= \lambda''/2.
\label{dcum}
\end{equation}
where the prime denotes derivatives taken at $z=0$. The explicit calculation of $\lambda(z)$ is usually quite difficult. 
However, it is possible to obtain $V$ and $D$ without calculating $\lambda(z)$ by using the following
method \cite{koza99}. The characteristic polynomial related to the matrix $\mathbf{L}(z)$ is defined as 
\begin{equation}
p(z,y)=\det\left(yI-\mathbf{L}(z)\right)= \sum_{q=0}^{Q}C_q(z)y^q,
\label{Pzy}
\end{equation}
where $C_q(z)$ are the coefficients of this characteristic polynomial. Since $\lambda(z)$ is a root of the polynomial, it follows that
\begin{equation}
\sum_{q=0}^{Q}C_q(z)\lambda^q(z)=0.
\label{chac1}
\end{equation} 
Moreover, for $z=0$ the matrix $\mathbf{L}(z)$ becomes the stochastic matrix, as given by Eqs. (\ref{matrix1}) and (\ref{matrix2}) for the two case studies, which has maximum eigenvalue $\lambda(0)=0$. Therefore
by taking the first derivative with respect to $z$ in Eq. (\ref{chac1}) we obtain
\begin{equation}
V=   -\frac{C_0'}{C_1},
\label{relaJ}
\end{equation}
and by taking the second derivative we get
\begin{equation}
D= -\frac{C_0''+2C_1'V+2C_2 V^2}{2C_1},
\label{relaD}
\end{equation}
where we have used Eqs. (\ref{Vcum}) and (\ref{dcum}). Hence by calculating these coefficients of the characteristic polynomial associated with
$\mathbf{L}(z)$ we obtain the Fano factor $F=2D/V$.  

%%%%%%%%%%%%%%%%%%%%%%%%%%%%%%%%%%%%%%%%%%%%%%%%%%%%%%%%%%%%%%%%%%%%%%%%%%%%%%%%%%%%%%%%%%%%%%%%%%%%%%%%%%%%%%%%%%%%%%%%%%%%%%%%%%%%%
%%%%%%%%%%%%%%%%%%%%%%%%%%%%%%%%%%%%%%%%%%%%%%%%%%%%%%%%%%%%%%%%%%%%%%%%%%%%%%%%%%%%%%%%%%%%%%%%%%%%%%%%%%%%%%%%%%%%%%%%%%%%%%%%%%%%%
\section{Explicit calculations for a simple network}
%%%%%%%%%%%%%%%%%%%%%%%%%%%%%%%%%%%%%%%%%%%%%%%%%%%%%%%%%%%%%%%%%%%%%%%%%%%%%%%%%%%%%%%%%%%%%%%%%%%%%%%%%%%%%%%%%%%%%%%%%%%%%%%%%%%%%
%%%%%%%%%%%%%%%%%%%%%%%%%%%%%%%%%%%%%%%%%%%%%%%%%%%%%%%%%%%%%%%%%%%%%%%%%%%%%%%%%%%%%%%%%%%%%%%%%%%%%%%%%%%%%%%%%%%%%%%%%%%%%%%%%%%%%
\label{appBnew}

\begin{figure}
\includegraphics[width=45mm]{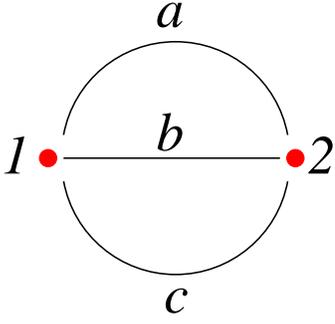}
\vspace{-2mm}
\caption{Two-state network used for the explicit calculations.}
\label{figBnew} 
\end{figure}

For illustrative purposes we calculate the Fano factor explicitly for the two-state network shown in Fig. \ref{figBnew}. We assume that both cycles
in this network have affinity
\begin{equation}
\A= \ln\frac{k_{12}^ak_{21}^b}{k_{21}^ak_{12}^b}=\ln\frac{k_{12}^ck_{21}^b}{k_{21}^ck_{12}^b}.
\label{affBnew}
\end{equation}  
Assuming that a product is generated in jumps from $1$ to $2$ through channels $a$ and $c$, the modified generator (\ref{modgenA}) for this model reads  
\begin{equation}
\left(
\begin{array}{cc}
-r_1 & k_{21}^a\textrm{e}^{-z}+k_{21}^b+k_{21}^c\textrm{e}^{-z}   \\
 k_{12}^a\textrm{e}^{z}+k_{12}^b+k_{12}^c\textrm{e}^{z}     & -r_2  \\ 
\end{array}
\right),  
\end{equation}  
where $r_1= k_{12}^a+k_{12}^b+k_{12}^c$ and $r_2= k_{21}^a+k_{21}^b+k_{21}^c$. The coefficients of the characteristic polynomial (\ref{Pzy})
associated with this matrix are $C_2(z)=1$, $C_1(z)=r_1+r_2$, and
\begin {equation}
C_0(z)=(\textrm{e}^{-z}-1)(\textrm{e}^{z}s_1 - s_2),
\end{equation}
where $s_1=(k_{12}^a + k_{12}^c) k_{21}^b$ and $s_2=(k_{21}^a + k_{21}^c) k_{12}^b$.
Using expressions (\ref{relaJ}), (\ref{relaD}) and $F=2D/J$ we obtain the following Fano factor,
\begin {equation}
F= \frac{s_1+s_2}{s_1-s_2}-2\frac{s_1-s_2}{(r_1+r_2)^2}.
\label{FBnew}
\end{equation}

As in the two case studies in the main text the bound (\ref{boundfinal})  is saturated when the transition rates are such that one cycle with uniform rates
dominates. This feature can be demonstrated by setting the rates as $k_{12}^a=k_{21}^b= \exp(\A/2)$, $k_{21}^a=k_{12}^b= 1$, $k_{12}^c= k\exp(\A/2)$, and 
$k_{21}^c= k$. With these rates, the Fano factor (\ref{FBnew}) becomes   
\begin{equation}
F= \frac{4\textrm{e}^{\A/2}(1 + k) + (\textrm{e}^{\A}+1)(2 + 2 k + k^2)}{(\textrm{e}^{\A}-1)(2+k)^2},
\end{equation}  
which is an increasing function of $k\ge 0$. The bound is saturated for $k\to 0$, where 
\begin{equation}
F= \frac{1}{2}\coth\left(\frac{\A}{4}\right).
\end{equation}

%%%%%%%%%%%%%%%%%%%%%%%%%%%%%%%%%%%%%%%%%%%%%%%%%%%%%%%%%%%%%%%%%%%%%%%%%%%%%%%%%%%%%%%%%%%%%%%%%%%%%%%%%%%%%%%%%%%%%%%%%%%%%%%%%%%%%
%%%%%%%%%%%%%%%%%%%%%%%%%%%%%%%%%%%%%%%%%%%%%%%%%%%%%%%%%%%%%%%%%%%%%%%%%%%%%%%%%%%%%%%%%%%%%%%%%%%%%%%%%%%%%%%%%%%%%%%%%%%%%%%%%%%%%
\section{Further examples of multicylcic networks}
%%%%%%%%%%%%%%%%%%%%%%%%%%%%%%%%%%%%%%%%%%%%%%%%%%%%%%%%%%%%%%%%%%%%%%%%%%%%%%%%%%%%%%%%%%%%%%%%%%%%%%%%%%%%%%%%%%%%%%%%%%%%%%%%%%%%%
%%%%%%%%%%%%%%%%%%%%%%%%%%%%%%%%%%%%%%%%%%%%%%%%%%%%%%%%%%%%%%%%%%%%%%%%%%%%%%%%%%%%%%%%%%%%%%%%%%%%%%%%%%%%%%%%%%%%%%%%%%%%%%%%%%%%%
\label{appB}

We have also tested our bound expressed in (\ref{boundfinal}) for other multicyclic networks. In this Appendix we discuss two examples that we have analyzed extensively.
For all cases discussed below we verified the bound (\ref{boundfinal}) with two independent checks: a random search in the space of transition rates and numerical minimization of the Fano factor.

\begin{figure}
\includegraphics[width=50mm]{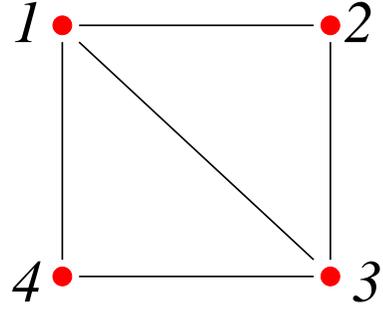}
\vspace{-2mm}
\caption{Further examples: four-state network of states.
}
\label{figB1} 
\end{figure}

The first example is the four-state network in Fig. \ref{figB1}. This network has three cycles: $\mathcal{C}_I= (1,2,3,1)$, $\mathcal{C}_{II}= (1,4,3,1)$, and $\mathcal{C}_{III}= (1,2,3,4,1)$. 
The affinities of the three-state cycles are $\A_I=\ln[k_{12}k_{23}k_{31}/(k_{21}k_{32}k_{13})]$ and $\A_{II}=\ln[k_{14}k_{43}k_{31}/(k_{41}k_{34}k_{13})]$. The affinity of the four-state cycle 
depends on the other two affinities as $\A_{III}= \A_I-\A_{II}$. The affinity of all cycles must be an integer multiplying $\A= \mu_s-\mu_p$, because of the physical constraint that there are 
only one substrate and one product. We have tested our bound for the following three cases. First, taking $\A_{I}=\A_{II}= \A$ and $\A_{III}=0$ we obtain $M_{\textrm{max}}=3$. For this choice 
the bound is saturated if one of the three-state cycles dominate. This case happens if, for example, we set $k_{12}=k_{23}=k_{31}= \exp(\A/3)$, $k_{21}=k_{32}=k_{13}= 1$, $k_{14}=k\exp(\A/3)$,
$k_{41}=1/k$, $k_{43}=\exp(\A/3)/k$, and $k_{34}=k$, with $k\to 0$. Second, taking $\A_{I}=\A_{III}= \A$ and $\A_{II}=0$ we obtain $M_{\textrm{max}}=4$. With this choice
the bound is saturated by setting the rates as  $k_{12}=k_{23}=k_{34}=k_{41}= \exp(\A/4)$, $k_{21}=k_{32}=k_{43}=k_{14}= 1$, $k_{13}=k\exp(\A/2)$, and $k_{13}=k$, with $k\to 0$. Third,
we have considered the case $\A_{I}=-\A_{II}= \A$ and $\A_{III}= 2\A$, for which $M_{\textrm{max}}=3$. The bound is saturated if one of three-state cycles dominate, which happens if we choose
the rates as $k_{12}=k_{23}=k_{31}= \exp(\A/3)$, $k_{21}=k_{32}=k_{13}= 1$, $k_{14}=k\exp(-2\A/3)$,
$k_{41}=1/k$, $k_{43}=\exp(-2\A/3)/k$, and $k_{34}=k$, with $k\to 0$.

\begin{figure}
\includegraphics[width=45mm]{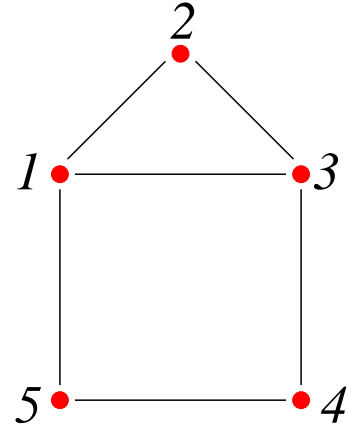}
\vspace{-2mm}
\caption{Further examples: five-state network of states.
}
\label{figB2} 
\end{figure}

The second example is the five-state network in Fig. \ref{figB2}. This network has three cycles: $\mathcal{C}_I= (1,2,3,1)$, $\mathcal{C}_{II}= (1,3,4,5,1)$, and $\mathcal{C}_{III}= (1,2,3,4,5,1)$.
Their affinities are $\A_I=\ln[k_{12}k_{23}k_{31}/(k_{21}k_{32}k_{13})]$,  $\A_{II}=\ln[k_{13}k_{34}k_{45}k_{51}/(k_{31}k_{43}k_{54}k_{15})]$, and $\A_{III}= \A_I+\A_{II}$. We have verified our
bound for the following two cases. First, with $\A_{I}=\A_{II}= \A$ and $\A_{III}=2\A$ we obtain $M_{\textrm{max}}=4$. The Fano factor saturates the bound if the four-state cycle $\mathcal{C}_{II}$ dominates, e.g.,
$k_{13}=k_{34}=k_{45}=k_{51}=\exp(\A/4)$, $k_{31}=k_{43}=k_{54}=k_{15}=1$, $k_{12}= k\exp(3\A/4)$, $k_{21}= k_{23}= 1/k$, and $k_{32}=k$, with $k\to 0$. Second, with 
$\A_{I}=8\A$ and $\A_{III}=-\A_{II}=4\A$ we obtain $M_{\textrm{max}}=5/4$. The bound is then saturated if $\mathcal{C}_{III}$ dominates, e.g., 
$k_{12}=k_{23}=k_{34}=k_{45}=k_{51}=\exp(4\A/5)$, $k_{21}=k_{32}=k_{43}=k_{54}=k_{15}=1$, $k_{31}= k\exp(32\A/5)$, $k_{13}= k$, with $k\to 0$.

\bibliography{refs}

\end{document}